\documentclass{PoS}

\def\deg{{$^{\circ}$}}
\def\bd17{\mbox{BD +17\deg 3248}}
\def\cs22{\mbox{CS 22892-052}}
\def\gtaprx{ \mathrel{  \vcenter{
                        \offinterlineskip \hbox{$>$}
                        \kern 0.3ex \hbox{$\sim$}    } } }

\title{In Memory of Al Cameron}

\ShortTitle{In Memory of Al Cameron}

\author{\speaker{John J. Cowan}\\
        Homer L. Dodge Dept. of Physics, University of Oklahoma
Norman, OK 73019\\
        E-mail: \email{cowan@nhn.ou.edu}}

\author{James W. Truran\\
        Department of Astronomy and
Astrophysics, University of Chicago,
Chicago, IL 60637 and 
Argonne National Laboratory, Argonne, IL 60439\\
        E-mail: \email{truran@nova.uchicago.edu}}

\abstract{
Al Cameron,
who died recently (October 3, 2005) at 80, 
was one of the giants in astrophysics. His insights 
were profound  and his interests were wide-ranging. 
Originally trained as a nuclear physicist,
he made major contributions in a number of fields,
including nuclear reactions in stars, nucleosynthesis,
the abundances of the elements in the Solar System, 
and the origin of the Solar System and the Moon.
In 1957, Cameron and, independently, Burbidge, Burbidge, Fowler 
and Hoyle, wrote seminal papers on nuclear astrophysics. 
Most of our current ideas concerning element formation in stars 
have followed from those two pioneering and historical works.
Al also made many contributions in the field of Solar System physics.
Particularly noteworthy in this regard was Cameron's work on the formation 
of the Moon. 
Al was also a good friend and mentor of young people. 
Al Cameron 
will be missed 
by many in the community both for his scientific contributions and for 
his friendship.
}

\FullConference{International Symposium on Nuclear Astrophysics --- Nuclei in the Cosmos --- IX\\
		 June 25-30 2006\\
		 CERN, Geneva, Switzerland}

\begin{document}

\section{Introduction}
Al Cameron was born June 21, 1925 in Winnipeg, Canada. 
During his long career he made many scientific contributions. 
He had enormous curiosity and was endlessly inventive and extremely 
prolific.
In addition to his scientific contributions,
Al Cameron  had a distinguished career in public service 
including his years as the Chair of the Space Sciences Board 
of the National Academy of Sciences from 1976 to 1982\cite{tru05,was05}.
Cameron was a member of the faculty of Harvard University from 
1973 until his retirement in 1999. He then joined 
the faculty of the Lunar and Planetary Laboratory of 
the University of Arizona, where he was a Senior Research Scientist. 

\section{Some of His Scientific Accomplishments}

\subsection{Element Synthesis}

The papers by Cameron\cite{cam57,cam57b} and Burbidge, Burbidge, 
Fowler and Hoyle\cite{bbfh57} substantially defined the field of 
nucleosynthesis as we understand it today. 
Originally  
trained in nuclear physics, Al's interest in nuclear astrophysics 
was triggered by the observational results of Merrill\cite{mer52},
who detected the presence of the radioactive element technetium 
in the atmospheres of red-giant stars - thereby indicating that
technetium actually synthesized in these types of stars.
This led Al to work on determining the sources of neutrons for this 
nucleosynthesis process in stars. He identified\cite{cam55} 
the two sources of neutrons - the $^{13}$C($\alpha$,n)$^{16}$O and
$^{22}$Ne($\alpha$,n)$^{25}$Mg reactions - which remain today the main
reactions driving slow neutron capture synthesis (the $s$-process) in red
giants and massive stars, respectively. 

Al contributed important 
discussions of carbon, neon, and oxygen thermonuclear reactions 
which provided critical early insights into the nature of the carbon, 
neon, and oxygen burning phases of core evolution of massive stars. Al 
also made enormous contributions to our understanding of the neutron 
capture processes responsible for the synthesis of most of the isotopes 
of nuclei in the mass range A > 70. The distinguishing characteristics 
of the $s$-process and $r$-process were contained in those early works. 
They also identified a likely - although still not confirmed - site 
for $r$-process nucleosynthesis, {\it i.e.}, in the neutron-rich material 
outside the core of an exploding Type II supernova. 
Al spent many years and developed many different models concerning the 
conditions and environments for the $r$-process\cite{tru78,cow83,cam83,cow85}. 
Even in his later years, Al continued to try and understand how
the $r$-process elements were formed and to identify the site for
this nucleosynthesis process\cite{cam01,cam03}.

\subsection{Explosive Nucleosynthesis of Iron Peak Nuclei}

Another significant product of Cameron's early intuition has been our
understanding of the character of explosive nucleosynthesis, of its
critical role in the production of the iron-peak nuclei present in nature,
and of the associated prediction that nuclei of mass A = 56 should be
formed predominantly as $^{56}$Ni in supernova events. Early efforts to
fit the iron abundance peak observed in nature were dominated by
attempts to provide the best fit to an iron (nuclear statistical equilibrium)
peak centered rather on $^{56}$Fe (see, {\it e.g.}, \cite{hoy60}).
Cameron recognized, however, that the burning timescale was the critical
factor: when nuclear burning proceeds on a hydrodynamic ({\it e.g.}, 
post shock)
timescale, weak interaction processes cannot act to convert any
appreciable fraction of protons to neutrons during the course of the
thermonuclear burning epoch. It follows that, for matter initially
dominated by self-conjugate nuclei ({\it e.g.} $^{12}$C, $^{16}$O, and
$^{28}$Si), the final products of explosive nucleosynthesis must also
lie along or very near to the Z = N line. Utilizing nuclear reaction
and weak interaction rates developed by Cameron and his students 
(described below),  
this
behavior was subsequently confirmed by detailed numerical calculations.
These calculations\cite{tru67}  
revealed $^{56}$Ni to be the most
abundant product of such explosive burning episodes with clear, and now
widely recognized, implications for the powering of the light curves of
supernovae.

\subsection{Solar System Abundances}

Cameron recognized the importance of a detailed understanding of the 
abundances of the elements in Solar System material as a guide to 
the nucleosynthetic processes by which they were produced. 
The abundance pattern itself confirmed the dominant role of nuclear 
processes in nucleosynthesis and displayed such regularities as the 
iron abundance peak (a nuclear statistical equilibrium feature) and 
the peaks in the heavy element region attributable to the effects of 
neutron shell features.  
His considerations of the neutron capture processes
ultimately led to the detailed separation and identification of the
distinct $s$-process and $r$-process isotopic components in the heavy element
region published in his discussion of the abundances of the 
elements\cite{cam59}.
Cameron's breakdown of the  
solar abundances into the $s$-process, $r$-process, and $p$-process
components, originally identified by 
Burbidge 
{\it et al.}\cite{bbfh57}, 
are shown in Figure 1\cite{cam73a}. 
It is significant to note that recent
studies of heavy element abundances in extremely metal deficient stars
(\cite{tru02,sne03a,cow06})
closely fit the $r$-process
pattern identified in Solar System matter by Cameron's early work. 
In fact our ideas on heavy element synthesis, and the assignments of the
Solar System abundances into their respective $r$, $s$ and $p$-components,
have deviated little from his in almost 50 years.
Cameron
and his collaborators also explored $r$-process synthesis in neutron-rich
supernova ejecta in early dynamic $r$-process calculations\cite{del71}.

\begin{figure}[ht]
\centering
\includegraphics[angle=0,height=3.50in]{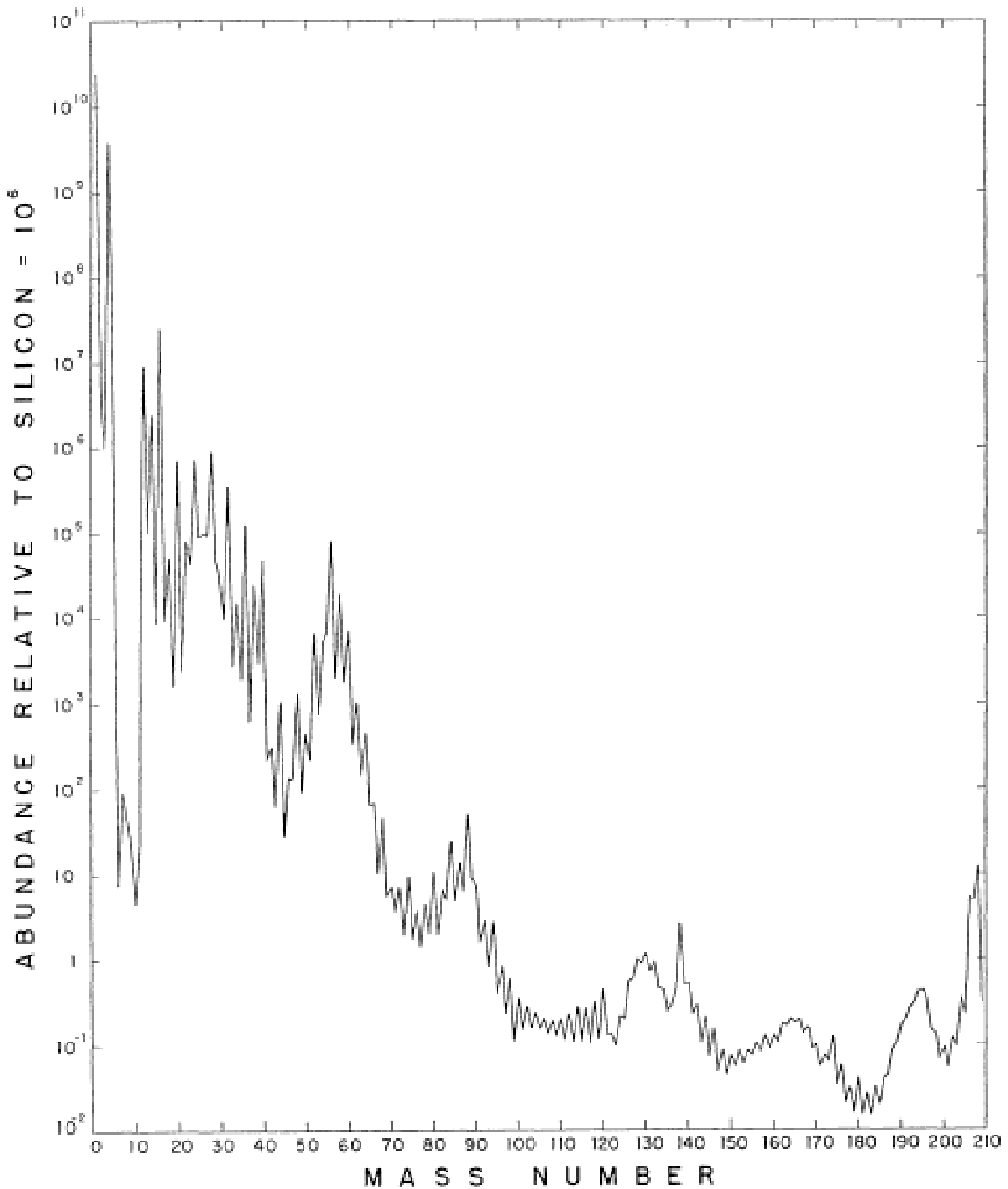}
\includegraphics[angle=0,height=3.60in]{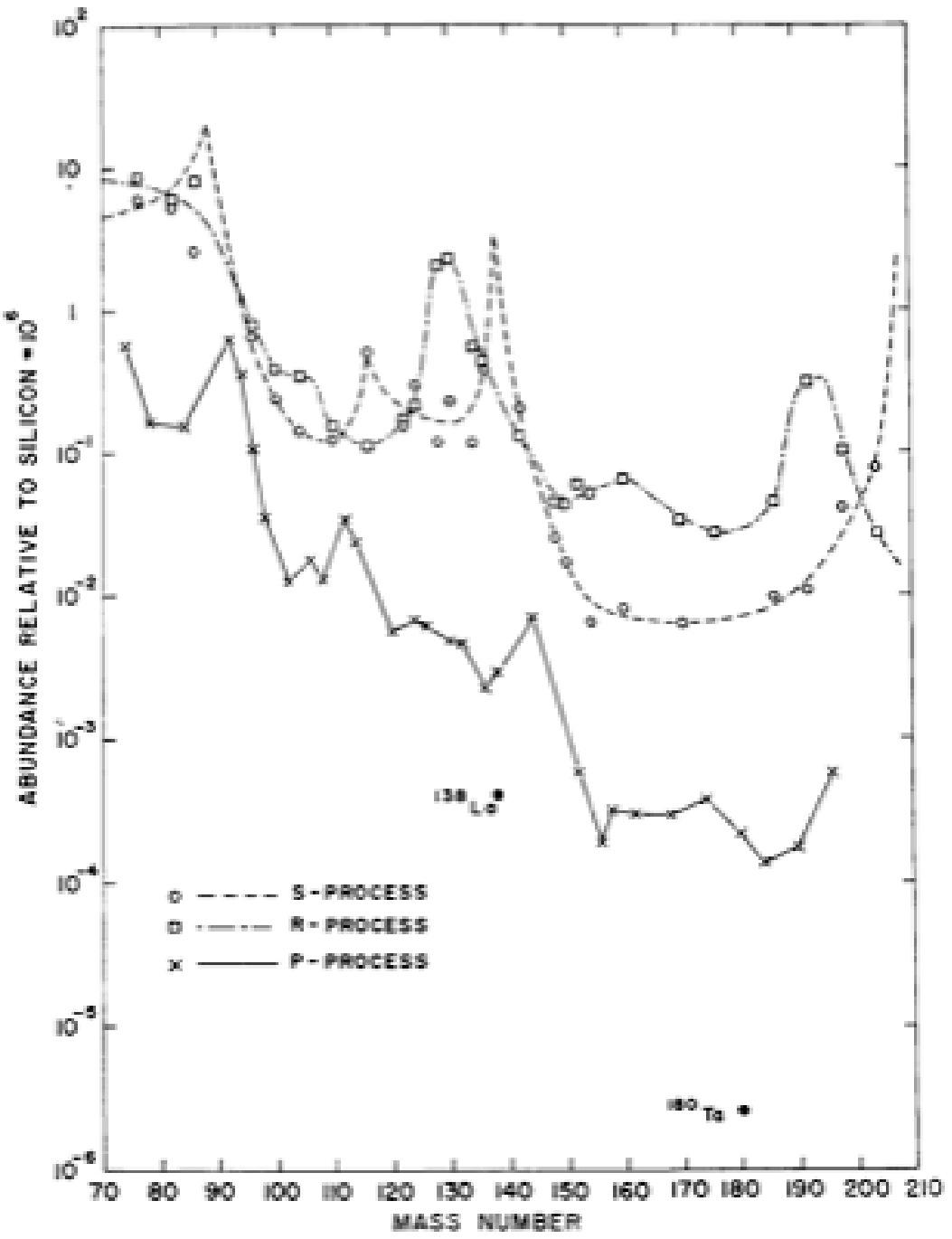}
\caption{
(left) Total and (right) components 
by nucleosynthetic process of the Solar System abundances from 
Cameron\cite{cam59,cam73a}.
\label{fig1}}
\end{figure}

\subsection{Nuclear Data for Astrophysics}

Cameron was also concerned with providing critical nuclear reaction rate 
and cross section data for for astrophysics calculations. 
He recognized that, while experimental studies in nuclear physics
can provide many of the nuclear reaction cross sections critical to the
early phases of stellar energy generation (e.g. hydrogen, helium, carbon, and
oxygen burning), this was not sufficient for studies of the later phases of
burning and nucleosynthesis, where hundreds of individual reactions can 
be involved. He pioneered the use of systematics of nuclear
properties that ultimately made possible theoretical calculations of heavy
element synthesis in diverse astronomical environments. Noteworthy in this
regard are studies by Cameron and his students 
of  nuclear mass formulae\cite{cam65},
of nuclear level densities\cite{cam58a,gil65},
of thermonuclear reaction rates\cite{tru66},
and of weak interaction rates\cite{han68}.
The rates arising from these early studies enabled critical and defining 
early investigations 
of explosive nucleosynthesis in supernovae. A clear measure of the impact
of this research in nuclear astrophysics is provided, for example, by the
fact that the study of nuclear level densities remains today the basis for
virtually all current calculations of thermonuclear reaction rates from
nuclear systematics.

\subsection{Solar System Studies and the Origin of the Moon}

Cameron also performed fundamental and pioneering work in studies of the
Solar System. He developed, for example, 
the idea of the supernova trigger for 
the formation of the Solar System\cite{cam77}. 
In this model, a single event could account for most isotopic anomalies 
and extinct radioactivities in the solar system material. 
He also spent many years trying to understand the formation of the Moon.
His work established the idea of a large  (Mars-sized) object 
impacting  on the proto-Earth, ejecting matter into orbit, where it 
accreted to form the Moon\cite{benz86}. As a result of years of numerical
simulations, on a large number of distributed computers, Al's calculations
demonstrated that this large-impact hypothesis 
could 
explain the masses, densities  and
angular momentum of the Earth-Moon system.

\begin{figure}[ht]
\centering
\includegraphics[angle=0,width=2.90in]{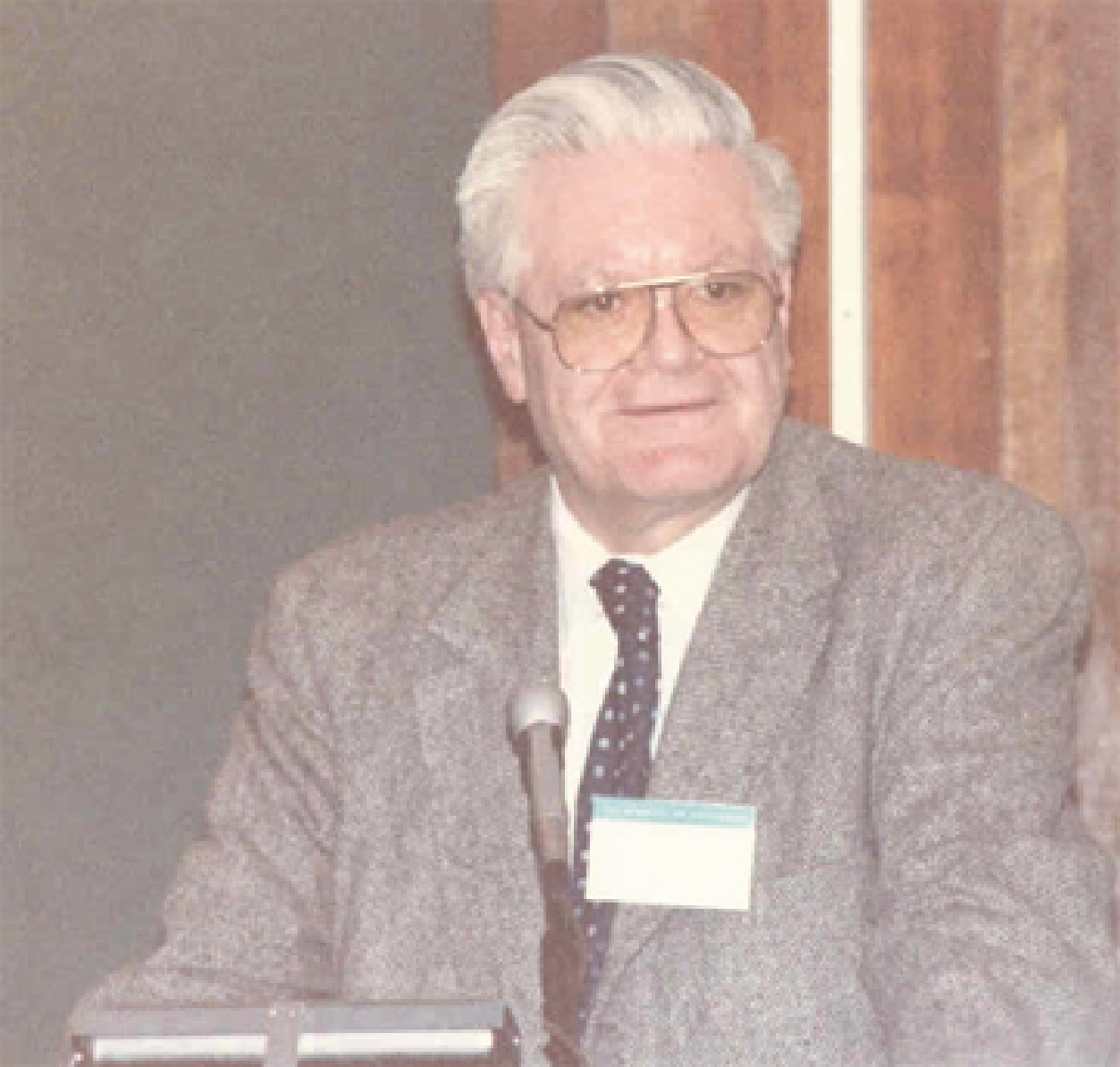}
\includegraphics[angle=0,width=2.90in]{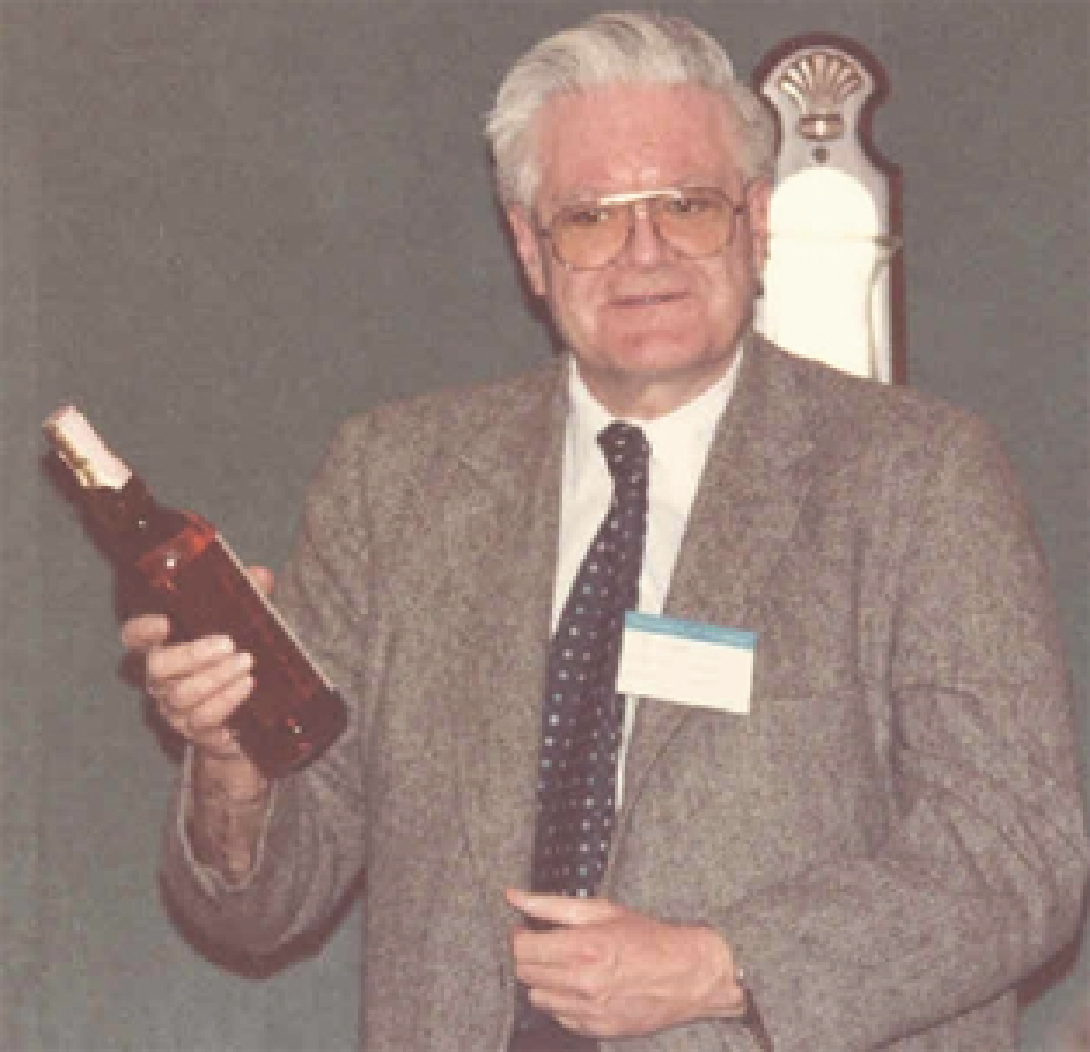}
\caption{
Al at work and play at a meeting in Boulder, CO in 1985 to celebrate
his 60th birthday (see \cite{cam86}).
\label{fig2}}
\end{figure}

\section{Legacies}

Al  left many scientific legacies: 
his work on nucleosynthesis, nuclear physics input for astrophysical 
environments, cosmic abundances and studies of the 
origin of the Solar System and
the Moon. 
Even more important 
Al was a good friend and mentor of young people.  
Al Cameron (1925-2005), who died in October of 2005, 
will be missed by many for his scientific curiosity and creativity,
his sense of humor, kindness and warmth.

\begin{figure}[ht]
\centering
\includegraphics[angle=0,width=3.90in]{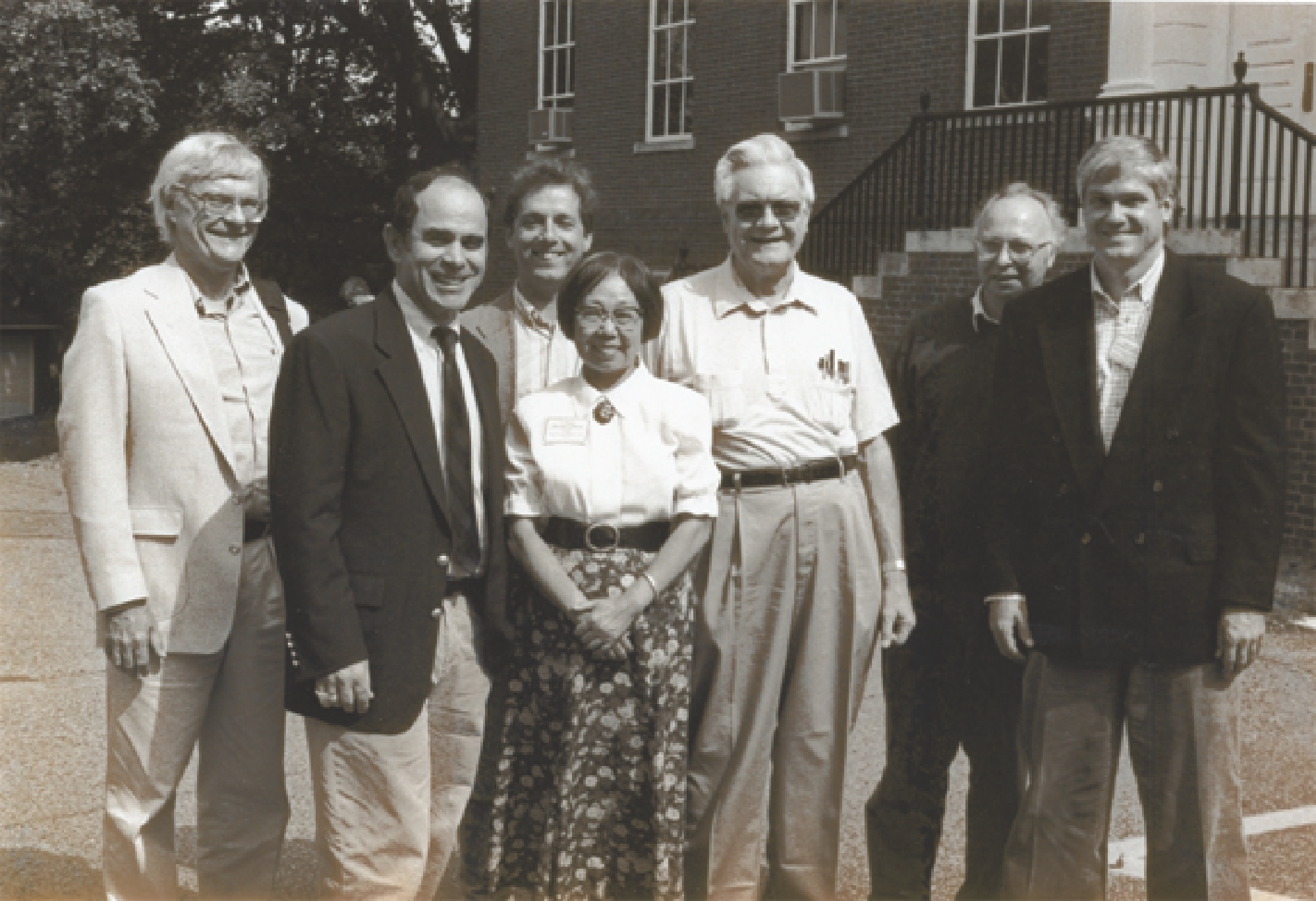}
\caption{ 
Al with some of his former students and postdocs in September of 
1995 to celebrate his 70th birthday at the Harvard-Smithsonian Center
for Astrophysics, in Cambridge, MA. 
\label{fig3}}
\end{figure}

\begin{figure}[ht]
\centering
\includegraphics[angle=0,width=\textwidth]{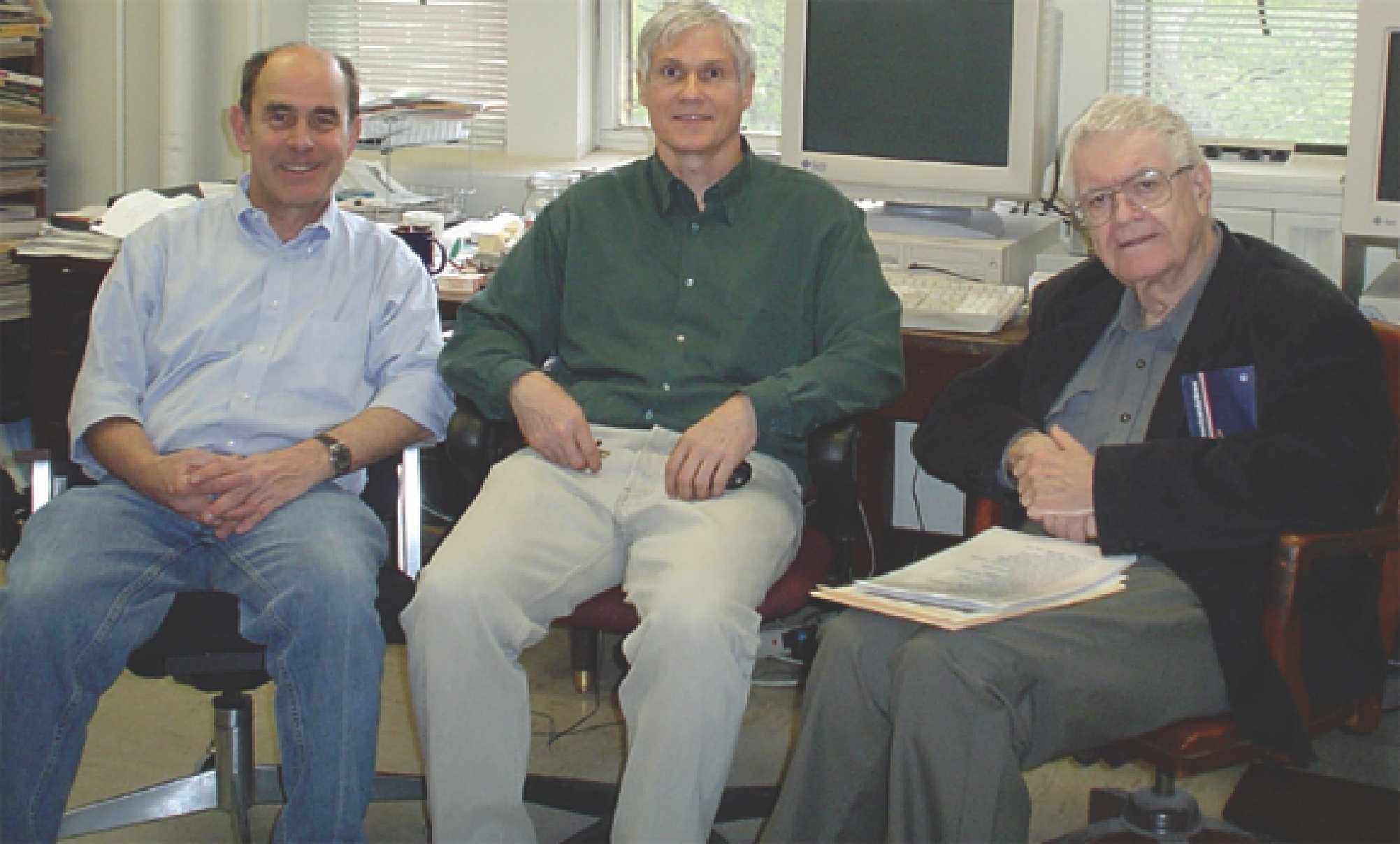}
\caption{
Al - still working on the $r$-process and just shortly before giving 
an 80 minute colloquium - with the authors in Norman, OK (April, 2005).
\label{fig4}}
\end{figure}

\begin{figure}[ht]
\centering
\includegraphics[angle=0,width=2.90in]{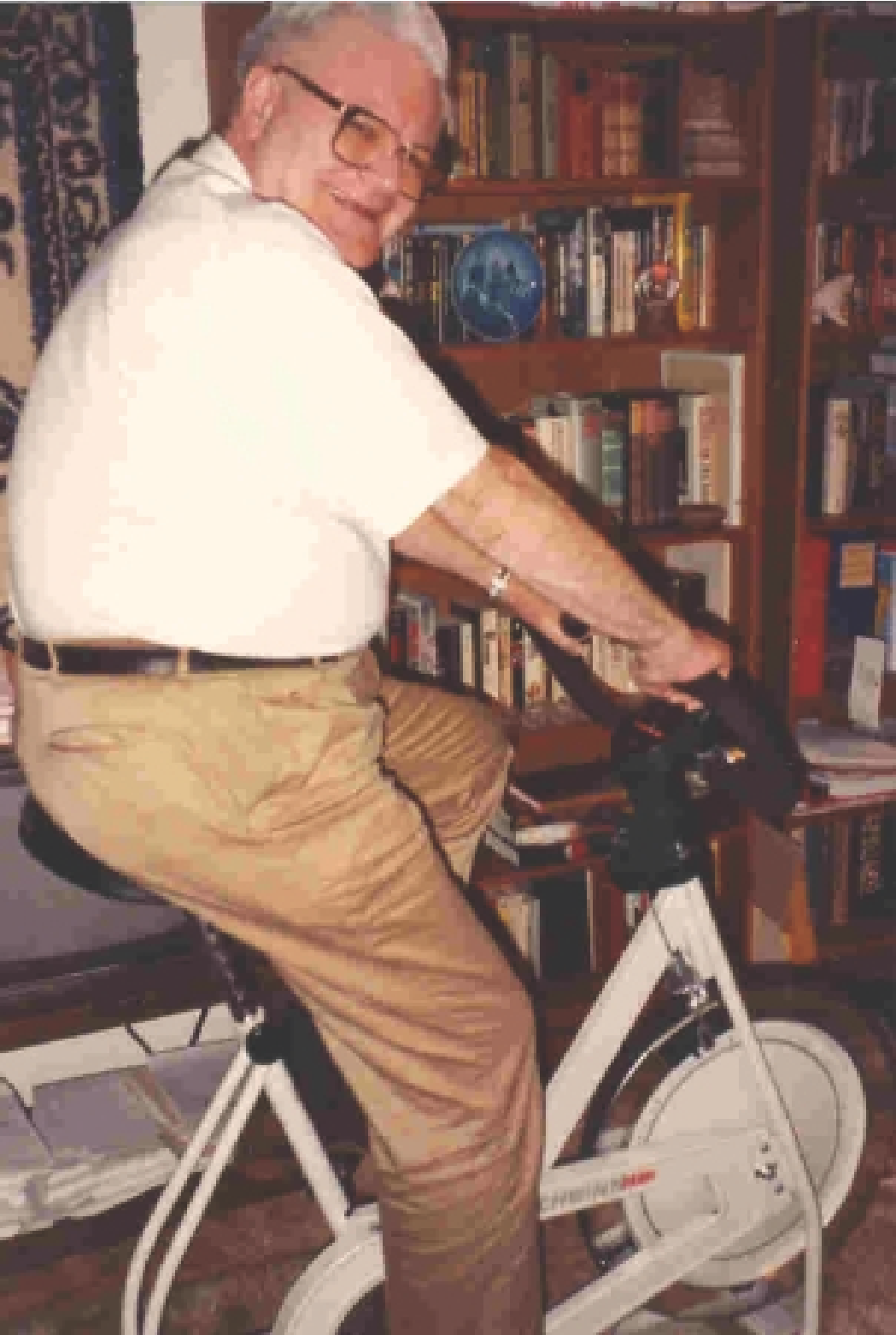}
\caption{
The only known picture of Al Cameron ever exercising taken in 
Norman, OK - showing Al's spirit 
of adventure and willingness to try anything, even exercise, at least
once. 
\label{fig5}}
\end{figure}

\acknowledgments
We thank C. Sneden for helpful comments.
The  work of the authors has been supported in part by the NSF, DOE 
and by STScI.

\end{document}